\documentstyle[eqsecnum,aps]{revtex}

\begin{document}

\title{Constraint on the magnetic moment of the top quark}
\author{R. Mart\'{\i}nez$^1$ and J.-Alexis Rodr\'{\i}guez$^{1,2}$ \\
1. Depto. de F\'{\i}sica, Universidad Nacional, Bogot\'a, Colombia  \\
2. Centro Internacional de F\'{\i}sica,  Bogot\'a, Colombia} 
\date{}
\maketitle

\begin{abstract}
We derive a bound on the magnetic dipole moment of the top quark in the context of
the effective Lagrangian approach by using the ratios
$R_b=\Gamma_b/\Gamma_{h}$, $R_l=\Gamma_{h}/\Gamma_l$ and the $Z$ width. We 
take into account the vertex and oblique corrections. 
\end{abstract}

The most recent analyses of precision measurements at the Large
Electron Positron (LEP) collider lead to the
conclusion that the predictions of the Standard Model (SM) of electroweak interactions,
based on
the gauge group $SU(2)_L\otimes U(1)_Y$ are in excellent agreement with
the experimental results. Recently the discovery of the top quark has been
announced by the Collider Detector at Fermilab (CDF) and D0 collaborations\cite{cdf}. The direct measurement of
the top quark mass $m_t$ is in agreement with the indirect estimates derived by
confronting the SM $m_t$ dependent higher order corrections with the LEP and
other experimental results. The measurement of the top quark mass reduced the number of
free parameters
of the SM. A precise knowledge of the value of the top mass will
improve the sensitivity of searches of new physics through small indirect
effects.

The precise measurements of the $g-2$ value of the electron provides a test of its
point-like character. Similarly, measurements of the electric- and chromo-
magnetic moments of the
quarks can be important to study physics beyond the
SM. In particular, the chromomagnetic moment of the top quark can affect its
production in the $p\overline{p}$ and $e^{+}e^{-}$ reactions\cite{pro}.

The SM predicts how the top quark should behave under these
interactions, so any deviation from this behaviour would provide us with a
probe of new physics beyond the SM. If new physics is found in this sector,
it could probably originates from a non standard symmetry breaking mechanism.
This is because the top mass is of the order of the electroweak (EW) breaking scale,
and hence it is conceivable that the top-quark properties are sensitive to
unsuppressed EW breaking effects\cite{peccei}.

The aim of the present work is to extract indirect information on the
magnetic dipole moment of the top quark from LEP data, specifically
we use the ratios $R_b$ and $R_l$ defined by 
\begin{eqnarray}
R_b&=&\frac{\Gamma (Z\rightarrow b\bar{b})}{\Gamma (Z\rightarrow hadron)}
\; , \nonumber \\
R_l&=&\frac{\Gamma (Z\rightarrow hadron)}{\Gamma (Z\rightarrow l\bar{l})}
\label{rbl} 
\end{eqnarray}
and the $Z$ width,
in the context of an effective Lagrangian approach. The oblique and QCD
corrections to the $b$ quark and hadronic $Z$ decay
widths cancel off in the ratio $R_b$. This property makes $R_b$
very sensitive to direct corrections to the $Zb\overline{b}$ vertex,
specially those involving the heavy top quark \cite{nardi}, while $\Gamma_Z$ and $R_l$ are more sensitive to oblique corrections.

The effective Lagrangian approach is a convenient model independent
parametrization of the low-energy effects of the new physics
that may show up at high energies\cite{wudka}. Effective Lagrangians, employed to study
processes at a typical energy scale $E$ can be written as a power series in 
$1/\Lambda$, where the scale $\Lambda$ is associated with the heavy particles
masses of the underlying theory\cite{burgues}. The coefficients of the different
terms in the effective Lagrangian arise from integrating out the heavy
degrees of freedom that are characteristic of a particular model for new
physics.

In order to define an effective Lagrangian it is necessary to specify the
symmetry and the particle content of the low-energy theory. In our case, we
require the effective Lagrangian to be $CP$-conserving, invariant under SM
symmetry $SU(2)_L\otimes U(1)_Y$, and to have as fundamental fields the
same ones appearing in the SM spectrum. Therefore we consider a Lagrangian
in the form
\begin{equation}
{\cal L}_{eff}={\cal L}_{SM}+\sum_{n}\alpha_n {\cal O}^n
\end{equation}
where the operators ${\cal O}^n$ are of dimension greater than
four. In the present work, we consider the following dimension six and $CP$-conserving
operators,
\begin{eqnarray}
O_{uW}^{ab} &=&\bar{Q}_L^a\sigma ^{\mu \nu }W_{\mu \nu }^i\tau ^i\tilde{\phi}%
U_R^b  \; ,\nonumber \\
O_{uB}^{ab} &=&\bar{Q}_L^a\sigma ^{\mu \nu }YB_{\mu \nu }\tilde{\phi}U_R^b \; ,
\label{operador}
\end{eqnarray}
where $Q_L^a$ is the quark isodoublet, $U_R^b$ is the up quark isosinglet,
$a$ , $b$ are the family indices,
$B_{\mu \nu }$ and $W_{\mu \nu }$ are the $U(1)_Y$ and $SU(2)_L$ field
strengths, respectively, and $\tilde{\phi}=i\tau_2\phi^*$. We use the
notation introduced by Buchm\"{u}ller and Wyler \cite{wyler}. In the case of
the operators $O_{uB}^{ab}$ and $O_{uW}^{ab}$, some degrees of family
mixing is
made explicit (corresponding to $a\neq b$) without breaking SM gauge invariance. After spontaneous
symmetry breaking, these fermionic operators generate also effective vertices
proportional to the anomalous magnetic moments of quarks. The above operators
for the third family give rise to the anomalous  $t\bar{t}\gamma $ vertex and
the unknown coefficients  $\epsilon _{uB^{}}^{ab}$ and
$\epsilon_{uW}^{ab}$ are related respectively with the anomalous magnetic
moment of the
top quark through 
\begin{equation}
\delta \kappa _t=-\sqrt{2}{\frac{m_t}{m_W}}\frac g{eQ_t}(s_W\epsilon
_{uW}^{33}+c_W\epsilon _{uB}^{33})\; .
\end{equation}
where $s_W$ denotes the sine of the weak mixing angle.

The expression for $R_b$ is given by 
\begin{equation}
R_b=R_b^{SM}(1+(1-R_b^{SM})\delta _b) \; ,
\end{equation}
where $R_b^{SM}$ is the value predicted by the SM and $\delta _b$ is the
factor which contains the new physics contribution, and it is defined as
follows 
\begin{equation}
\delta _b=2{\frac{{g_V^{SM}g_V^{NP}+g_A^{SM}g_A^{NP}}}{{(g_V^{SM})^2+(g_A^{SM})^2}}}
\end{equation}
and $g_{V}^{SM}$ and $g_A^{SM}$ are the vector and axial vector couplings of the $Zb\bar{b}$
vertex normalized as $g_V^{SM}=-1/2+2 s_W^2/3$ and $g_A^{SM}=-1/2$. The
contributions from new physics, eq. (\ref{operador}), to $R_l$ and
$\Gamma_Z$ are of two classes. One from vertex correction to $Zb\bar{b}$
in the $\Gamma_{hadr}$ and the other from the oblique correction through
$\Delta\kappa$ in the $\sin^2\theta_W$. These can be written as
\begin{eqnarray}
R_l&=&R_l^{SM} (1-0.1851\; \Delta\kappa +0.2157 \;\delta_b)   \; ,\nonumber \\
\Gamma_Z&=&\Gamma_Z^{SM} (1-0.2351\;\Delta\kappa+0.1506 \;\delta_b)
\end{eqnarray}
where $\Delta\rho$ is equal to zero for the operators that we are considering.

The contribution of the above effective operators to the 
$Zb\overline{b}$ vertex is given by the Feynman diagrams shown in fig. 1,
where a heavy dot denotes an effective vertex. After evaluating the Feynman
diagrams, with insertions of the effective operators $O_{uB}^{ab}$ and
$O_{uW}^{ab}$ we obtain 
\begin{eqnarray}
g_V^{NP} &=&4 \sqrt{2} \epsilon _{uW}^{33} G_F m_W^3 m_t \big \{
3 c_W (\tilde{C}_{12}-
\tilde{C}_{11})-{\frac{m_t^2}{\sqrt{2} m_W^2}}(C_{12}-C_{11}+C_0)
\nonumber \\
&&+{\frac{(1+a)}{8c_W}}(C_{11}+C_{12}+C_0)+\frac{1}{\sqrt{2}}
(C_{12}-C_{11}-C_0) \nonumber \\
&&-\frac{3a}{4c_Wm_Z^2}(B_1-B_0)%
\big \} \; ,\\
g_A^{NP} &=&4 \sqrt{2}\epsilon _{uW}^{33} G_f m_W^3 m_t \big \{
-{\frac a{2c_W}}%
(C_0+C_{12}-C_{11})-\frac{1}{\sqrt{2}}(C_{12}-C_0-C_{11})
\nonumber \\
&&+\frac{m_t^2}{\sqrt{2} m_W^2}(C_{12}-C_{11}+C_0)
-{\frac{2 m_ts_W^2}{m_W}}(\tilde{C}_0+\tilde{C}_{12}-%
\tilde{C}_{11})  \nonumber \\
&&-\frac{3}{4c_Wm_Z2}(B_1-B_0)\big \}  \; 
\end{eqnarray}
for the operator $O_{uW}$ and, 
\begin{eqnarray}
g_V^{NP} &=&g_A^{NP}=\frac{4\sqrt{2}}{3}\epsilon _{uB}^{33} G_F m_W^3 m_t
\frac{s_W}{c_W} [\frac{%
m_t^2}{\sqrt{2}m_W^2}(C_{12}-C_{11}+C_0)  \nonumber \\
&&-\frac{1}{\sqrt{2}}(-C_{11}+C_{12}-C_0)]  \; ,
\end{eqnarray}
for the operator $O_{uB}$. In the above equations $a=1-{\frac 83}s_W^2$ while $C_{ij}=C_{ij}(m_W,m_t,m_t)$,
$\tilde{C}_{ij}=\tilde{C}_{ij}(m_t,m_W,m_W)$ and $B_i=B_i(0,m_t,m_W)$ are the Passarino-Veltman scalar
integral functions \cite{veltman}. The combination $B_0-B_1$ has a pole in
$d=4$ dimensions that is identified with the logarithmic dependence on
the cutoff. Using the prescription given in ref. \cite{hariwara}, the pole can be
replaced by $\ln\Lambda^2/m_Z^2$.

The operators (\ref{operador}) contribute to the fermion processes at one
loop level, giving oblique corrections to the gauge boson self energies.
The contribution is essentially coming from the $\Sigma_{\gamma Z}(m_Z^2)$
self energy. Therefore these operators only contribute to $\Delta\kappa$
parameter \cite{oblicuas}. For $\Delta\kappa$ we have obtained the same
results of the eqs. (50) and (51) of ref. \cite{renard}. 

To obtain the physical quantities $R_b$, $R_l$ and $\Gamma_Z$ as a 
function of $\delta\kappa_t$ instead of two parameters 
$\epsilon^{33}_{uW}$ and $\epsilon^{33}_{uB}$, we consider that only one 
coefficient at the time is different from zero at the scale
$\Lambda$. 
Then we proceed to sum the value of the contribution of each operator 
in order to avoid cancellation between them. With this prescription we 
get an optimal bound.
 
In  Fig. 2 we display $R_b$ as a function of $\delta \kappa _t$.
The horizontal lines represent the experimental measurements
$R_b^{exp}=0.2178\pm 0.0011$ \cite{lep}; our bound can be expressed as 
$0.38\!\leq\! \delta\kappa_t\!\leq\! 1.21$. In Fig. 3 we show $R_l$ fraction
versus $\delta\kappa_t$ with the horizontal lines representing the
experimental result, $R_l^{exp}=20.778\pm 0.029$; for this case the bound 
can be expresses as $0.02\!\leq\!\delta\kappa_t\! \leq\! 0.48$. Finally in 
Fig. 4 we plot $\Gamma_Z$ versus $\delta\kappa_t$ with the experimental value
$\Gamma_Z^{exp}=2.4946\pm 0.0027$ GeV; the limit is  
$0\!\leq\!\delta\kappa_t\!\leq \! 0.48$. The SM values for the parameters that we have used are $\Gamma_Z=2.4972$
GeV, $R_l=20.747$ GeV, $R_b=0.2157$, $\Gamma_{hadr}=1743.4$ MeV and
$\Gamma_l=84.03$ MeV; with the input parameters: $m_t=175$ GeV,
$\alpha_s(m_Z)=0.118$, $m_Z=91.1861$ GeV, $m_H=100$ GeV and $\Lambda=1$ TeV.

In conclusion, the corrections through $R_l$ and $\Gamma_Z$ put a
better bound on $\delta\kappa_t$ than the vertex correction from $R_b$. Our results from figs. 3 and 4 are of the same order of the results of the eqs. (56) and (57) of 
ref. \cite{renard}, with the appropriate replacement between $f_{tW\phi}$ and $f_{tB\phi}$ and the magnetic dipole moment of the top quark. These bounds agrees also with
the one obtained of the same effective operators in reference \cite{martinez},
by using the CLEO result on $B(b\to s\gamma)$.

We would like to thank M. Perez and E. Nardi for their comments. We thank
COLCIENCIAS for financial support.

\newpage

\begin{center}
Figure Captions
\end{center}

\noindent Figure 1. Feynman diagrams contributing to the $Z\to b\bar{b}$ decay. 
The heavy dots denote an effective vertex.

\vspace{1cm}

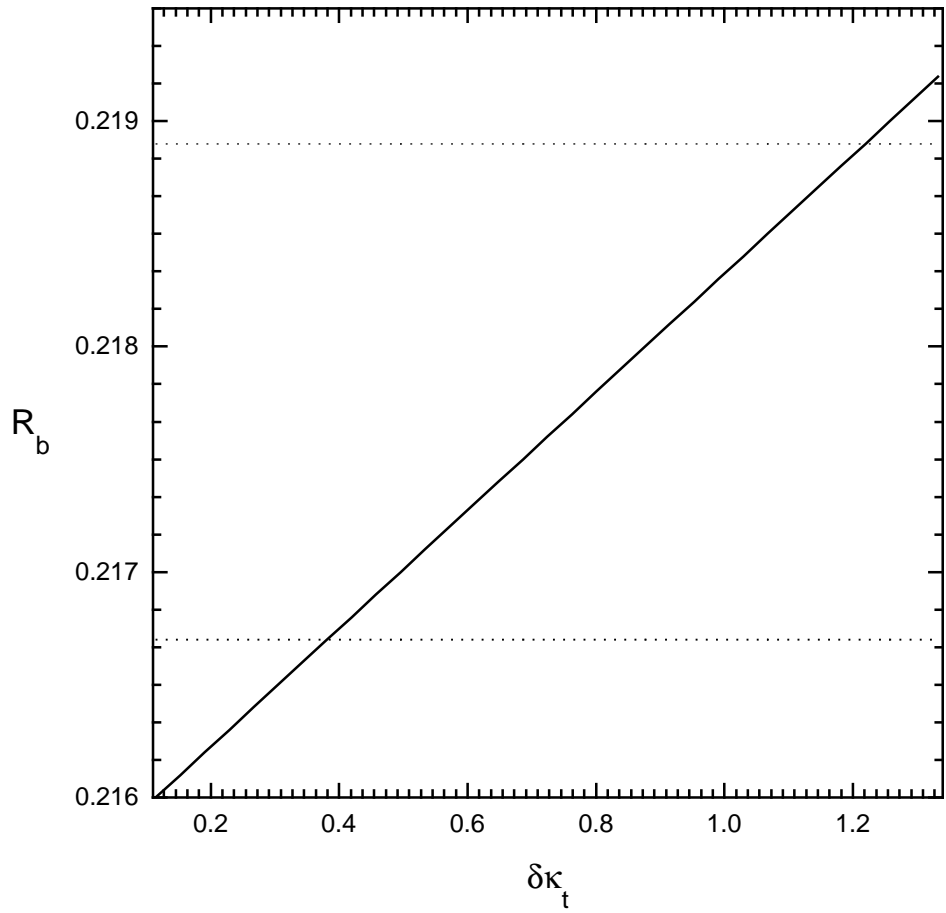
\noindent Figure 2. $R_b$ as a function of $\delta\kappa_t$ for $m_t=175$
GeV. The horizontal lines are the experimental results. 
\vspace{1cm}

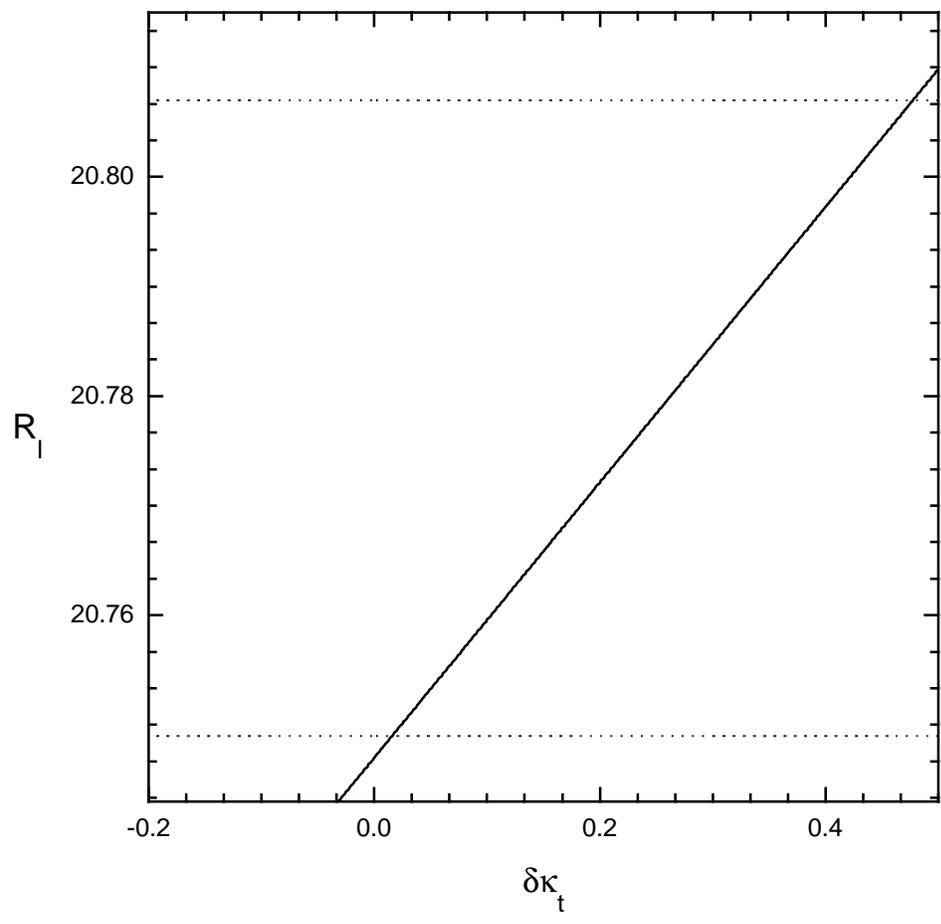
\noindent Figure 3. Same as fig. 2 for $R_l$.
 
\vspace{1 cm}

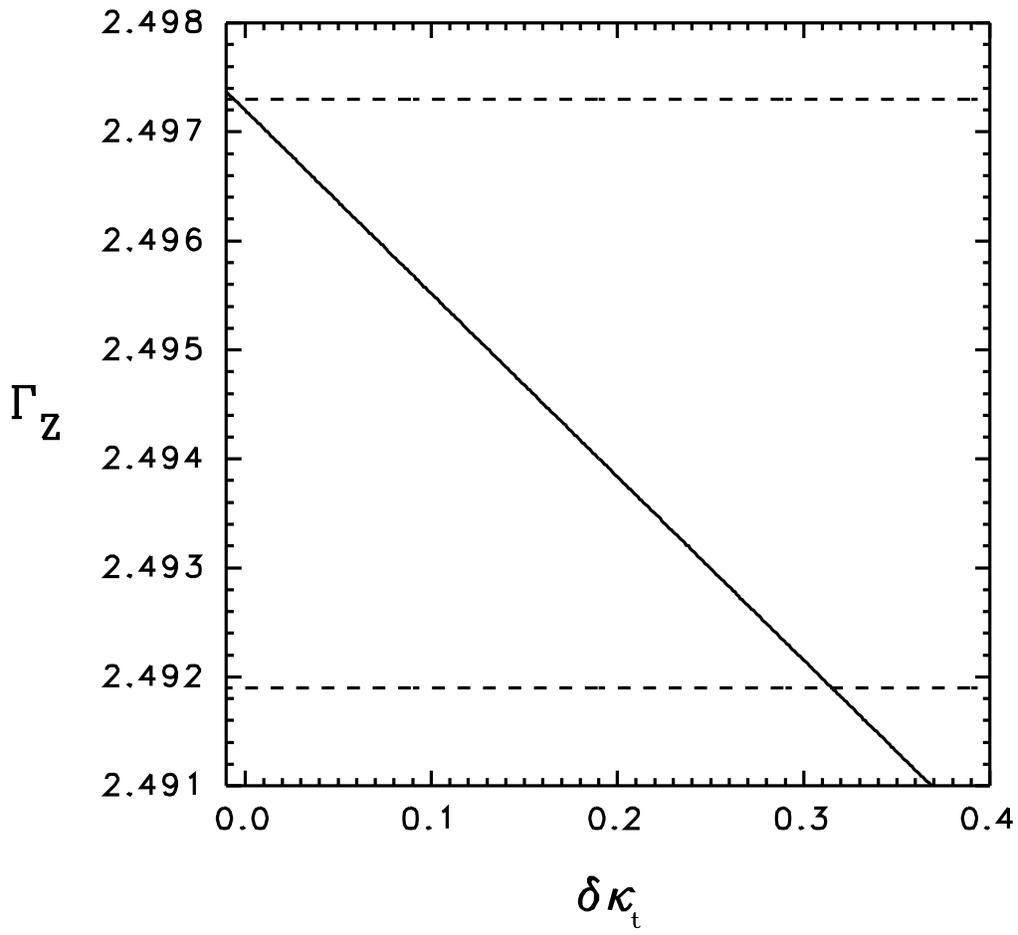
\noindent Figure 4. $\Gamma_Z$ as a function of $\delta\kappa_t$ for
$m_t=175$
GeV. The horizontal lines are the experimental results.

\end{document}